\documentclass[seceq,twocolumn]{jpsj2}
\title{Visualization of Superfluid Helium Flow}

\author{\textsc{Matthew S. Paoletti},$^{1,2}$ \textsc{Ralph B. Fiorito},$^1$ \textsc{Katepalli R. Sreenivasan}$^{3,4}$ and \textsc{Daniel P. Lathrop}$^{1,2,3,5}$\thanks{e-mail: lathrop@umd.edu}}

\inst{$^{1}$Institute for Research in Electronics and Applied Physics \\
$^{2}$Department of Physics \\
$^{3}$Institute for Physical Sciences and Technology \\
University of Maryland, College Park, MD, USA 20742\\
$^{4}$International Centre for Theoretical Physics, Trieste, Italy 34014\\
$^{5}$Department of Geology, University of Maryland, College Park, MD, USA 20742}

\abst{We discuss an experimental technique to visualize the motion in the bulk of superfluid $^4$He by tracking micron-sized solid hydrogen tracers.  The behavior of the tracers is complex since they may be trapped by the quantized vortices while also interacting with the normal fluid via Stokes drag.  We discuss the mechanism by which tracers may be trapped by quantized vortices as well as the dependencies on hydrogen volume fraction, temperature, and flow properties.  We apply this technique to study the dynamics of a thermal counterflow.  Our observations serve as a direct confirmation of the two-fluid model as well as a quantitative test of the normal fluid velocity dependence on the applied heat flux.
}

\kword{superfluid, quantum fluids, thermal counterflow, quantum turbulence, flow visualization}
\begin{document}
\maketitle

\newpage

\section{Introduction} \label{introduction} %% No sections necessary for express letters, letters and short notes
Superfluid $^4$He has received a great deal of recent attention owing to the newfound ability to image micron-sized tracer particles in the bulk of the fluid.  Polymer micro-spheres and hydrogen isotopes have been used for particle image velocimetry (PIV) measurements \cite{zhang04, zhang05a, zhang05b, vansciver07, xu07} and solid hydrogen tracers have been used to visualize the structure and dynamics of quantized vortices \cite{bewley06, bewley07, bewley08, paoletti08}.  However, the dynamics of tracers in superfluid flows are more complicated than in viscous fluids since the particles can interact with the quantized vortices \cite{parks66} in addition to the normal fluid.  The details of these interactions depend upon the temperature, particle characteristics, line-length density of quantized vortices, and flow properties.  Progress has been made in trying to understand precisely what tracer particles \lq \lq track" in superfluid turbulence \cite{poole05, sergeev06, barenghi07b, kivotides07, kivotides08}, though many issues remain unanswered.

In this review, we discuss a technique whereby micron-sized solid hydrogen tracers are imaged in the bulk of superfluid $^4$He.  Previous studies have shown that these tracer particles can be trapped by the quantized vortices \cite{bewley06}, thereby allowing us a direct visualization of the vortex dynamics, including reconnection \cite{bewley08, paoletti08}.  Even when the particles are trapped on vortices, they experience Stokes drag with the normal fluid, as they would with any viscous liquid.  The particle trapping mechanism and its dependencies are discussed in \S\ref{trapping} and \S\ref{dependencies}, respectively.  The effects of the hydrogen particles on the superfluid dynamics are mentioned in \S\ref{particle_effects}.  We discuss experiments using hydrogen tracers in counterflow turbulence in \S\ref{counterflow} and conclude in \S\ref{conclusions}.

\section{Particle Trapping Mechanism} \label{trapping}

The trapping of ions and particles by quantized vortices in superfluid $^4$He occurs by
a mechanism proposed by Parks and Donnelly in 1966 \cite{parks66}.  The superfluid motion around a locally straight vortex can be
expressed in cylindrical coordinates $\{s,\phi,z\}$ as $v_{\phi} = \kappa/2\pi s$ where $\kappa=h/m=9.97\times 10^{-4}$~cm$^2$/s is the quantum of circulation with $h$ as Planck's constant and $m$ the mass of a $^4$He atom.  Trapping a particle or bubble on the vortex core causes a reduction
in the kinetic energy owing to the displaced circulating superfluid helium. The reduction
in energy is maximized when the particle is centered on the vortex core (where the
kinetic energy density is largest).  As a particle approaches a vortex, a gradient in energy
causes an attractive force -- although some dissipative mechanism is required to prevent the particle from oscillating radially.
Drag between the particle and the normal fluid likely serves to dissipate
the energy as the particle approaches the vortex.

The attractive force between a tracer particle and a quantized vortex
can equivalently be described by pressure gradients.  The Bernoulli pressure around the vortex resulting from the superfluid
motion is $P = -\rho_\mathrm{s} \kappa^2 / 8 \pi^2 s^2$, where $\rho_\mathrm{s}$ is the superfluid density and $\kappa$ the quantum of circulation.  The normal force on the particle is obtained by integrating the pressure gradient on the surface of the particle.  If the particle is many particle diameters away from the vortex core
the total force may be approximated as $\mathbf{F}_\mathrm{trap} = \frac{4}{3} \pi a^3 \nabla P \sim -s^{-3}\hat{s}$.  Thus a local pressure gradient attracts particles on to the vortex, dissipatively mediated by Stokes drag.

This trapping mechanism is different than for a viscous vortex.  Since superfluid flow has zero viscosity, the particle is not dragged into orbit around the vortex.  In a viscous fluid, particles orbit azimuthally producing a centrifugal force in addition to the radial pressure gradient.  If a small particle is neutrally buoyant then the centrifugal force is balanced by the pressure gradient and the particle circulates the vortex at constant cylindrical radius.  Particles that are less dense than the fluid would be radially drawn to the vortex core whereas denser particles would be expelled to the boundaries.  In a superfluid, all particles, independent of density are drawn to the quantized vortices.  We have observed that solid hydrogen (less dense than liquid helium) and solid neon (denser than liquid helium) may be trapped by the quantized vortices.

\section{Particle Trapping Dependencies} \label{dependencies}

The forces that act on a tracer particle in a fluid may be broken into two categories -- those acting on the volume of the particle and normal surface forces.  The only force acting on the volume in our experiments is gravity.  The density of the solid hydrogen particles is slightly less than that of the liquid helium and the particles are therefore slightly buoyant.  The buoyancy force scales with the particle volume, whereas normal surface stresses scale with the surface area.  Therefore, keeping all else constant, the ratio of the surface forces to buoyancy increases with decreasing particle size.  Since we use constant density for all of our solid hydrogen tracer particles, smaller particles are more desirable since the effects of buoyancy are reduced relative to dynamic pressure and viscous forces.

The surface stresses acting on the particle are very important to the dynamics.  Specifically, the two most important forces in our experiments are Stokes drag induced by the normal fluid flow and trapping forces produced by the quantized vortices.  The interplay of these two determines whether the particles simply follow the normal fluid or get trapped or scattered by the quantized vortices.

As mentioned above, when the particle is far from the vortex the trapping force may be approximated by $\mathbf{F}_\mathrm{trap} = \frac{4}{3} \pi a^3 \nabla P \sim -s^{-3}\hat{s}$, where $a$ is the particle radius, $P=-\rho_\mathrm{s} \kappa^2 / 8 \pi^2 s^2$ the pressure, and $s$ the cylindrical radius from the vortex core.  $P$ is a linear function of the superfluid density $\rho_\mathrm{s}$, which depends nonlinearly on temperature.  The superfluid density is a small fraction of the total near the $\lambda$-transition, thereby making the trapping force weak.

Viscous drag is essential to the particle trapping mechanism, as mentioned earlier, since it provides a means of dissipation as the particle falls into the vortex core.  However, Stokes drag may also serve a different role.
Vortex motion transverse to the normal fluid may dislodge particles from the
vortices since Stokes drag is still relevant.  Therefore, at any temperature, two regimes always exist: (1) for sufficiently low relative velocities between the normal fluid and the quantized vortex the trapping force dominates and the particles remain trapped and (2) for sufficiently high relative velocities, Stokes drag is able to dislodge the particles from the vortex cores.  The resulting picture is that understanding the population of trapped and untrapped particles must necessarily
depend on both temperature and relative flows between the normal fluid and the quantized vortices.

\begin{figure}
\begin{center}
\includegraphics[width=8.6cm]{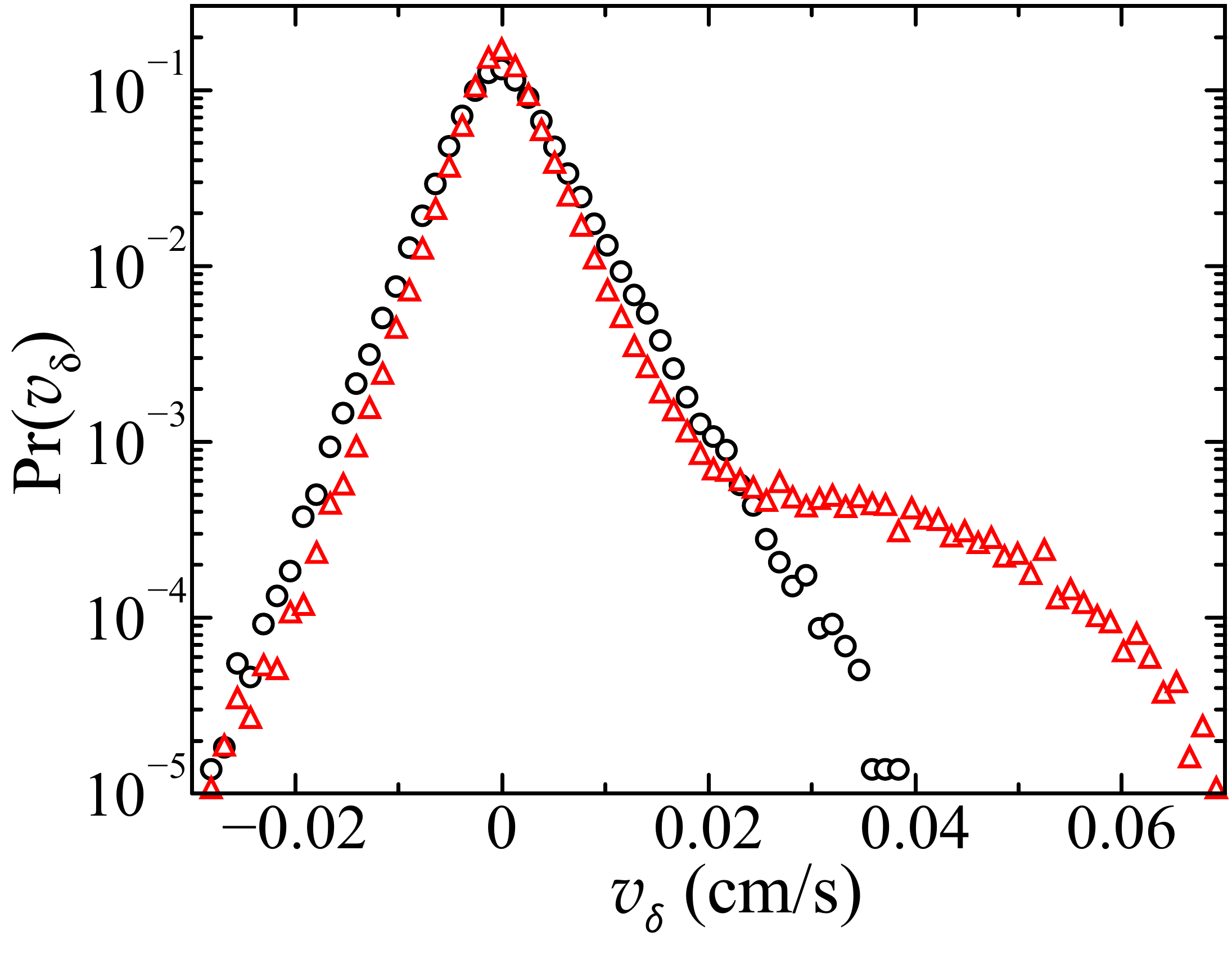}%
\caption{Statistics of pairwise particle separation velocities $v_\delta$.  The relative velocity between a pair of particles initially separated by less than 150~$\mu$m is determined by fitting the pairwise separation $\delta_{ij}(t)\equiv \left |\mathbf{r}_i(t)-\mathbf{r}_j(t) \right |$ to the form $\delta_{ij}(t)=v_\delta t+\delta_{ij}(0)$, where $\mathbf{r}_i(t)$ is the two-dimensional position vector of particle $i$ at time $t$. The distributions are computed from particle trajectories while the system is cooling at a rate of -152~$\mu$K/s with a 4~mW/cm$^2$ heat flux applied to drive a gentle counterflow between the normal fluid and superfluid.  Each data set spans 20 seconds with temperature values $51<T_\lambda-T<53$~mK for the black circles and $55<T_\lambda-T<57$~mK for the red triangles.  Only the lower temperature data (red triangles) show clear signs of counterflowing particles, evidenced by the tail for large $v_\delta$, which signifies strong drag forces.  We attribute this stark distinction to a crossover between Stokes-drag-dominated behavior at higher temperatures to a regime where particles are able to remain trapped on the quantized vortices.}
\label{vdelta}
\end{center}
\end{figure}

The competition between weak trapping forces near the $\lambda$-transition and normal fluid drag is evidenced in our experiments by the lack of trapped particles within $\sim$~50~mK of transition.  Trapped particles are often characterized by velocity vectors that clearly differ from the background normal fluid flow, thereby experiencing a strong Stokes drag.  To explore this temperature dependence we observe the particle dynamics as the system continues cooling below the $\lambda$-transition with a weak counterflow imposed to drive the vortices in opposition to the background flow (see \S\ref{counterflow} for more details on counterflows).  The details of a typical experiment are shown in Fig.\ \ref{vdelta}.  The result of these experiments is that there is a clear crossover from a regime where Stokes drag dominates to one where particles are able to become trapped on the quantized vortices.

In addition to temperature and flow properties, the competition between Stokes drag and particle trapping also depends upon particle size.  Stokes drag scales linearly with the radius of the particle.  Since the particle trapping force arises from excluded kinetic energy of circulating superfluid, the trapping force can be thought of as depending on the average kinetic energy density of the excluded superfluid.  Since the kinetic energy density is greatest near the vortex core, the average kinetic energy density excluded by a trapped particle increases with decreasing particle size.  Therefore, smaller particles, as we have observed, are more likely to become trapped by the quantized vortices. The technique of generating very small hydrogen particles is partly responsible for our ability to make the present observations.  Ions and electron bubbles would be even more greatly trapped as evidenced by the historical and current work using these systems.

\section{Particle Effects on Superfluid Dynamics} \label{particle_effects}

The hydrogen particles used in our experiments are not completely passive tracers.  By trapping a particle on the core of a vortex, the vortex also becomes trapped on the particle.  Thus, a particle trapped on a quantized vortex increases the coupling between the vortex and the normal fluid through the action of Stokes drag acting on the particle.  Quantized vortex trapping on the hydrogen particles is analogous to the pinning of dislocations and domain walls on impurities in a solid.  Realizing that the hydrogen particles may modify the stability and dynamics of the vortices we have endeavored to reduce the particle size and concentration.  In the limit of very few, small particles, one might expect to obtain a minor modification to the superfluid state.

The effects of the hydrogen tracers on the helium dynamics may be studied by observing the system as one modifies the volume fraction of hydrogen.  While a systematic variation of the volume fraction remains a useful future project, this section conveys our
general observations on volume fraction effects over the last several years.
While we give approximate volume fraction values below, the phenomenology
likely depends upon the particle size distribution, the quantized
vortex population (line-length density), and temperature.

\begin{figure}[tb]
\begin{center}
\includegraphics[width=4.6cm]{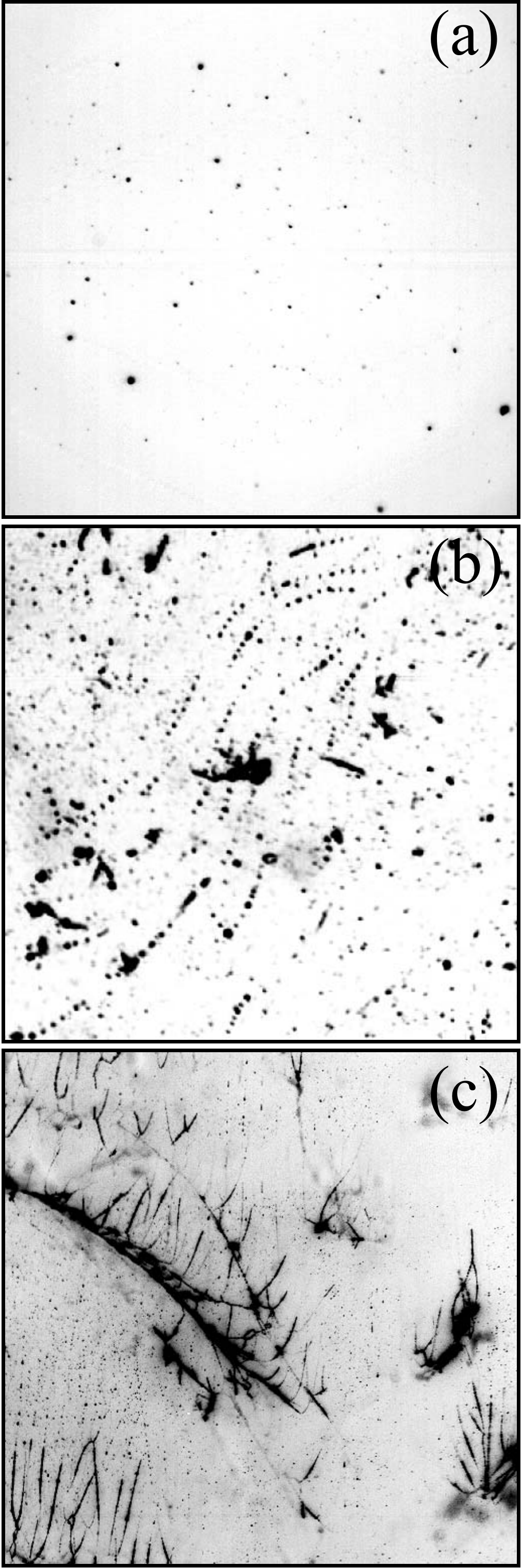}%
\caption{Intensity-inverted images showing varying hydrogen volume fractions.  The images in (a) and (c) show the full field of view (8~mm $\times$ 8~mm) and the image in (b) is 6.78~mm $\times$ 6.78~mm.  (a) The lowest volume fraction ($\phi_{\mathrm{H_2}}/\phi_{\mathrm{He}} < 10^{-8}$) used in our experiments.  (b) A moderate volume fraction $10^{-8}<\phi_{\mathrm{H_2}}/\phi_{\mathrm{He}} < 10^{-6}$), which often results in multiple particles trapped on each visible vortex.  In such cases, the trapped particles tend to be uniformly spaced along the vortex core.  (c) The more extreme effects of very large hydrogen volume fractions ($\phi_{\mathrm{H_2}}/\phi_{\mathrm{He}} >10^{-5}$) here in a rotating container.  By completely decorating the core of a quantized vortex, the hydrogen may serve to stabilize branches and networks of vortices that should be unstable in their absence.}
\label{volume_fraction}
\end{center}
\end{figure}

The smallest volume fractions ($\phi_{\mathrm{H_2}}/\phi_{\mathrm{He}} < 10^{-8}$)
are realized as only tens of micron-sized solid hydrogen particles in our field of view (8~mm~$\times$~8~mm~$\times$~100~$\mu$m) as shown in Fig.\ \ref{volume_fraction}(a).  The motions of the particles in superfluid helium, depending upon the various effects discussed above, split into two classes.  Many of the hydrogen
particles smoothly drift via Stokes drag with the normal fluid.  A subset of
the observed particles, though, have velocities that clearly differ from this smooth, background velocity field.  At times, the motions of closely spaced particles can even be antiparallel.  We interpret the second observation as suggesting the trapping of particles by the quantized vortices.

At moderate volume fractions ($10^{-8}<\phi_{\mathrm{H_2}}/\phi_{\mathrm{He}} < 10^{-6}$) we observe
multiple particles on each evident quantized vortex, e.g. Fig.\ \ref{volume_fraction}(b). Unexpectedly, the particles
are often uniformly spaced along the vortex core.  These
dotted lines appear with spacings that are both uniform on one vortex and
do not substantially vary between vortices \cite{bewleythesis}.

The observations of the uniform spacing imply particle-particle interactions present
when they are trapped on the quantized vortices. We might hypothesize that the interaction
originates from an inter-particle potential with both short-range repulsion
(at least hard-core), and a long-range attraction such that the balance
sets an equilibrium spacing (see Fig.\ \ref{volume_fraction}(b)).  The particle spacing is typically $\sim 100$~$\mu$m, which is much larger than the particle size.  These observations imply that the effective interaction between trapped particles differs from the bare interaction present in the bulk of the fluid.  Systematic studies of this feature should be performed.

At yet higher volume fractions of hydrogen ($\phi_{\mathrm{H_2}}/\phi_{\mathrm{He}} >10^{-5}$), the non-passivity
of the hydrogen particles is clearly evident as seen in Fig.\ \ref{volume_fraction}(c).  Often, the particles come to
fully cover the cores of the vortices.  In cases where a continuous cylinder of solid hydrogen is present on the vortex core,
branches and networks of vortices are stabilized, which should not be the case in the absence of particle loading.

\section{Thermal Counterflows} \label{counterflow}
\subsection{Two-fluid model}
Quantum fluids are often described as a mixture of two interpenetrating fluids \cite{landau41, tisza47}, a viscous normal fluid and an inviscid superfluid characterized by long-range quantum order.  The density and velocity of the normal fluid are denoted $\rho_{\mathrm{n}}$ and $\mathbf{v}_{\mathrm{n}}$ while the superfluid density and velocity denoted $\rho_{\mathrm{s}}$ and $\mathbf{v}_{\mathrm{s}}$.  The conservation of mass is given by $\rho_{\mathrm{n}} \mathbf{v}_{\mathrm{n}}+\rho_{\mathrm{s}} \mathbf{v}_{\mathrm{s}}=0$.  In a superfluid, entropy and heat are carried only by the normal fluid component.  Therefore, if a heat flux $\mathbf{q}$ is applied to the closed end of a channel in a superfluid then the normal fluid velocity will be aligned with the heat flux while the superfluid velocity will align oppositely to conserve mass.  Specifically, the simplest resulting normal and superfluid velocities for a given heat flux $\mathbf{q}=q\hat{z}$ are
\begin{eqnarray}
\mathbf{v}_{\mathrm{n}}=v_{\mathrm{n}} \hat{z}=\frac{q}{\rho S T} \hat{z}, \label{vn}\\
\mathbf{v}_{\mathrm{s}}=v_{\mathrm{s}} \hat{z}=-\frac{\rho_{\mathrm{n}}}{\rho_{\mathrm{s}}}v_{\mathrm{n}} \hat{z}, \label{vs}
\end{eqnarray}
where $S$ is the specific entropy, $T$ is the temperature, and $\rho=\rho_\mathrm{n}+\rho_\mathrm{s}$ is the density \cite{landau}.

\subsection{Previous studies}
Zhang and Van Sciver used solid polymer micro-spheres as tracers in counterflow turbulence \cite{zhang05a}.  The experiments were performed over the temperature range 1.62~K~$<T<2.0$~K for heat fluxes from 110 to 1370~mW/cm$^2$.  The velocities of the particles were measured using the particle image velocimetry (PIV) technique.  The most striking result from these experiments was that the velocities of the particles were approximately one half the average velocity of the normal fluid, independent of temperature.  This disparity was attributed to momentum transfer between the tracer particles and the quantized vortices.  Specifically, Zhang and Van Sciver proposed a form for a body force that would act on the tracer particles, which was shown to produce a temperature-independent velocity difference between $\mathbf{v}_\mathrm{n}$ and the observed particle velocity in agreement with their observations.  The results of these experiments were discussed theoretically by Sergeev \textit{et al} \cite{sergeev06}.  By applying theoretical arguments, they were able to obtain quantitative agreement with the experiments but at odds with eq.~(\ref{vn}).

\begin{figure}[tb]
\begin{center}
\includegraphics[width=8.6cm]{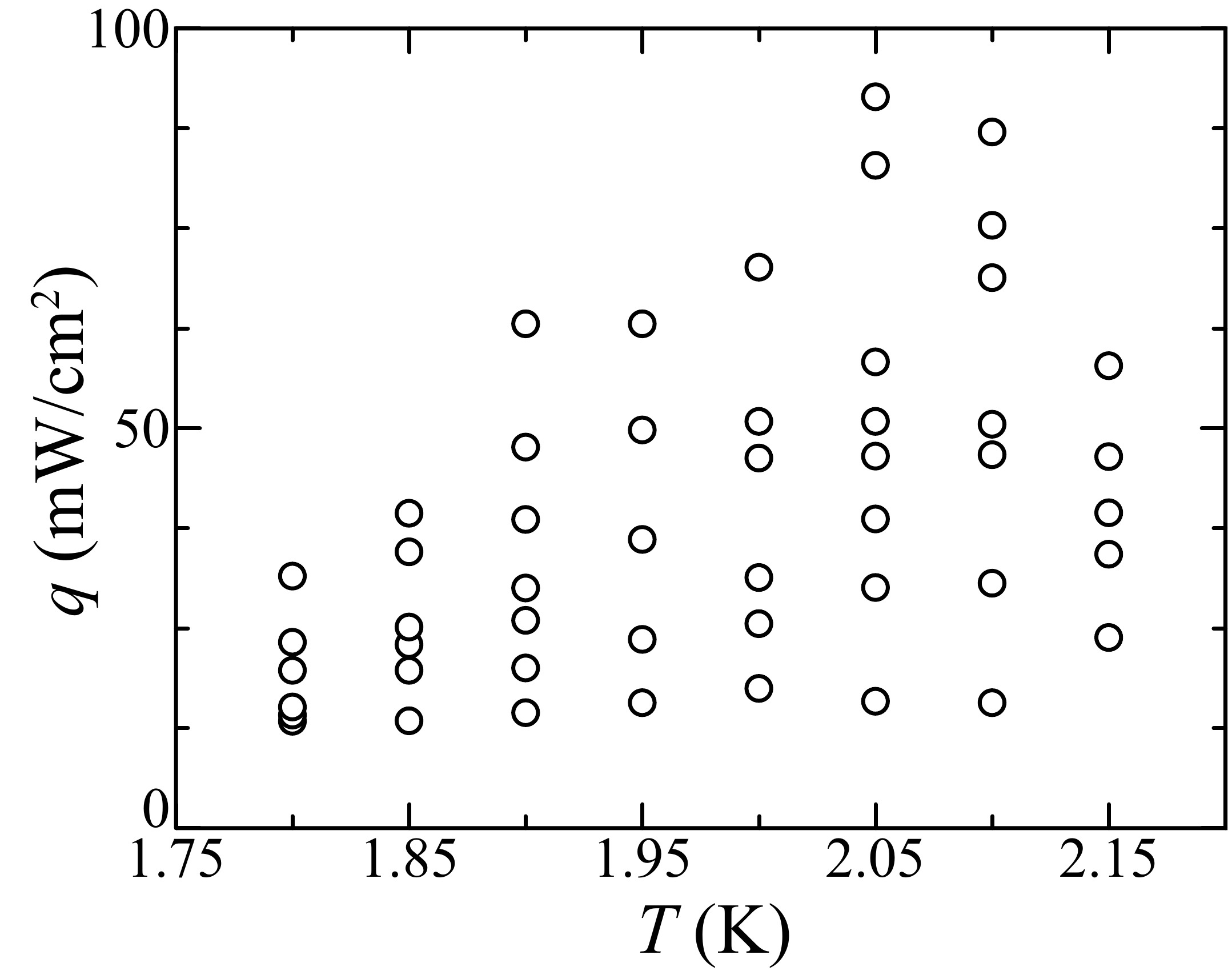}%
\caption{Parameter space diagram summarizing the thermal counterflow experiments.}
\label{param_space}
\end{center}
\end{figure}

\subsection{Present counterflow experiments}
The thermal counterflow experiments presented here are conducted in a cylindrical cryostat of 4.5~cm diameter containing liquid $^4$He.  The long axis of the channel is vertical with four 1.5~cm windows separated by 90$^{\circ}$.  A room-temperature mixture of 2\% H$_2$ and 98\% $^4$He is injected into the liquid helium above the superfluid transition temperature (2.17~K) producing a polydisperse distribution of tracer particles with diameters of order 1~$\mu$m \cite{bewley06}.  The initial volume fraction of hydrogen is approximately $10^{-6}<\phi_\mathrm{H_2}/\phi_\mathrm{He}<10^{-5}$.  The fluid is then evaporatively cooled to the desired temperature in the range 1.80 K $<T<2.15$~K.  A portion of the hydrogen leaves the observation volume resulting in a volume fraction $\sim 10^{-7}$.  The hydrogen particles are illuminated by a laser sheet that is 3.1~mm tall and 150~$\mu$m wide produced by a laser pointer.  Optical laser power is less than 5~mW.  A microchannel amplified CCD camera gathers 90$^{\circ}$ scattered light with a resolution of 24~$\mu$m per pixel at 50 frames per second.  A spiral nichrome wire heater located at the bottom of the channel 7.5~cm below the observation volume generates the counterflow by applying a fixed heat flux $\mathbf{q}=q\hat{z}$ with 13~$<q<91$~mW/cm$^2$.  Single particle trajectories are obtained using a two-dimensional particle-tracking algorithm with sub-pixel precision \cite{weeks}.  The experimental parameters used in these studies are summarized in Fig.\ \ref{param_space}.

\begin{figure}[b]
\begin{center}
\includegraphics[width=8.6cm]{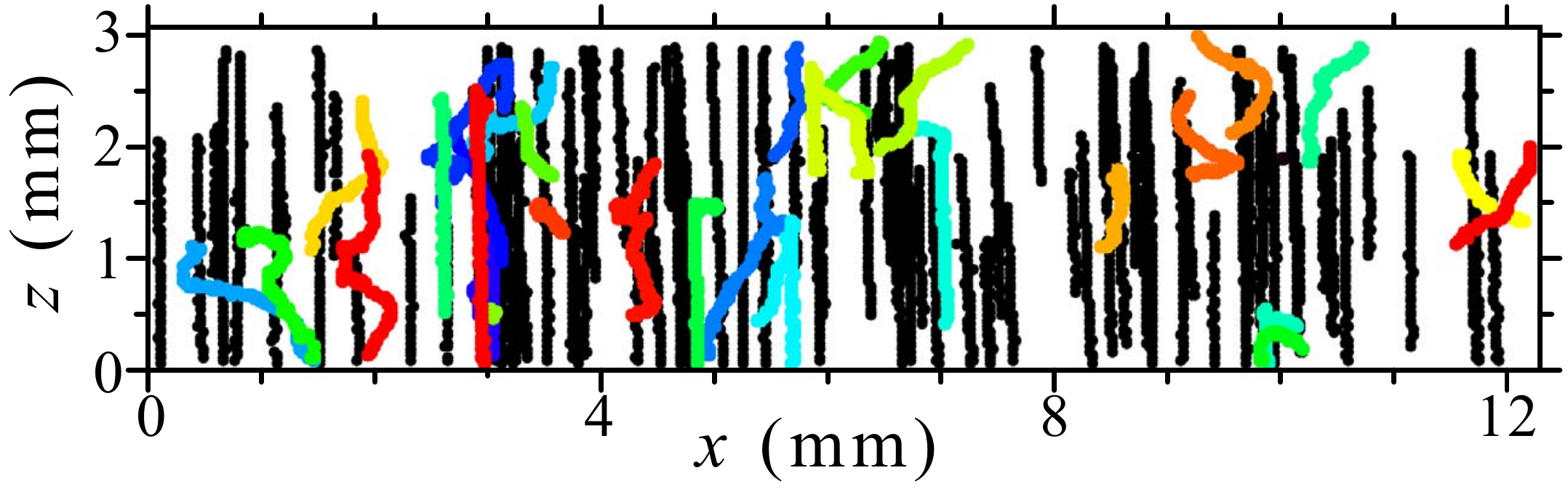}%
\caption{Example particle trajectories from a counterflow with $T=1.95$~K and $q=36$~mW/cm$^2$.  Trajectories that move upward with the normal fluid ($v_z>0$) are shown in black while tracers trapped in the vortex tangle ($v_z<0$) move downward and are shown in color.}
\label{cf_traj}
\end{center}
\end{figure}

We characterize the resulting dynamics by analyzing the particle trajectories.  Several example trajectories from a typical counterflow experiment are shown in Fig.\ \ref{cf_traj}.  Two distinct types of behavior are observed: (1) trajectories that move upward in the direction of the heat flux ($\hat{z}$) are denoted by black and (2) trajectories that oppose this motion and move downward ($-\hat{z}$) are shown in color.  The upward trajectories appear smooth and remarkably uniform, whereas the downward trajectories can be quite erratic.  In the context of the two-fluid model, we interpret upward-moving particles as being dragged by the normal fluid while downward-moving particles as trapped in the vortex tangle.

\subsection{Results}
\begin{figure}[tb]
\begin{center}
\includegraphics[width=8.6cm]{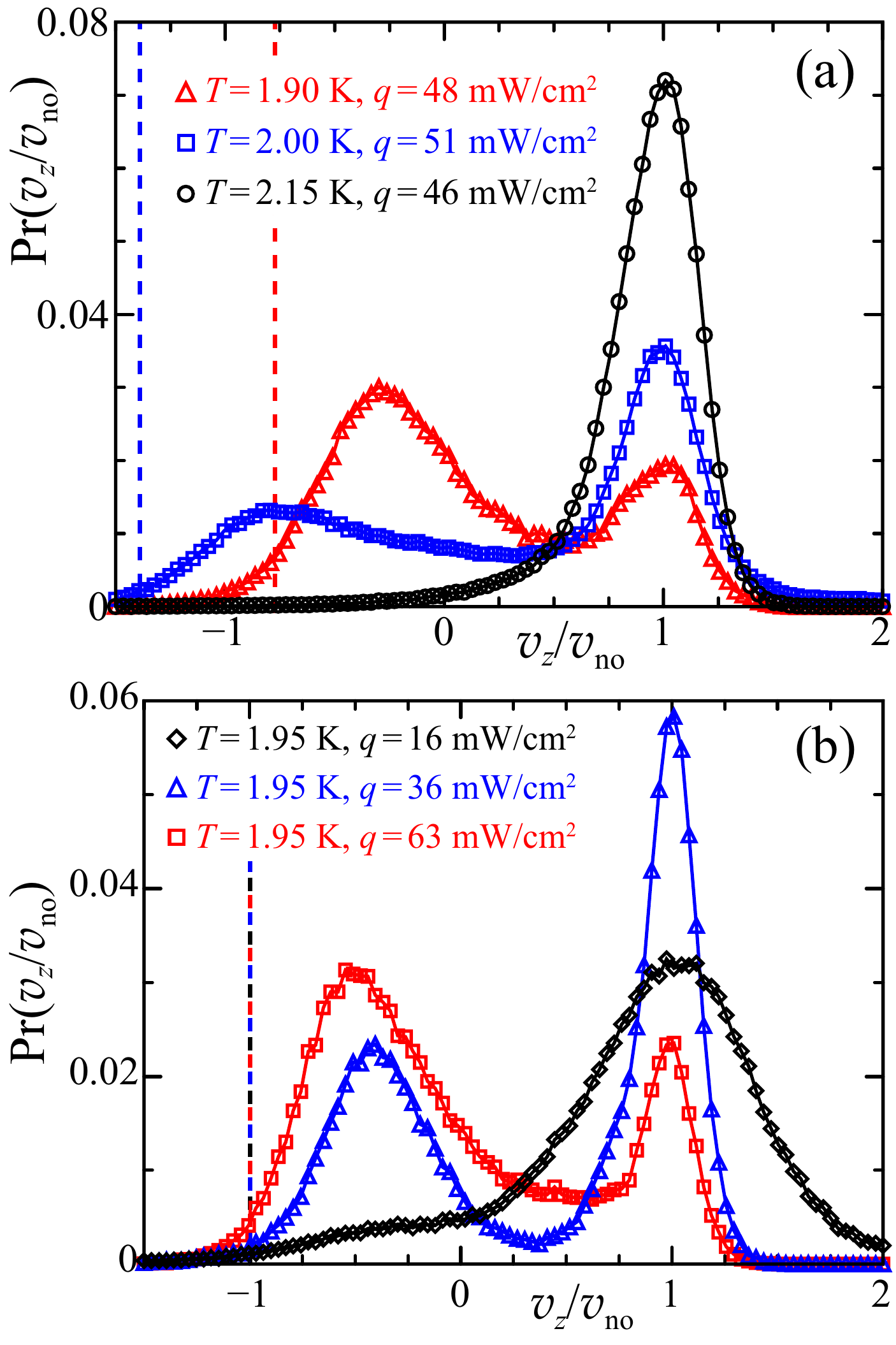}%
\caption{Vertical velocity statistics of example thermal counterflow experiments.  All of the vertical velocity $v_z$ distributions are scaled by the observed normal fluid velocity $v_{\mathrm{no}}$ as given in Fig.\ \ref{vplots}(a).  The predicted values of the superfluid velocity $v_\mathrm{s}$ given by eq.~(\ref{vs}) and scaled by $v_\mathrm{no}$ are shown by vertical dashed lines.  (a) Variation of vertical velocity distributions with varying temperature for (approximately) constant heat flux with the experimental parameters given by the legend.  (b) Variation of vertical velocity distributions with varying heat flux at constant temperature with the experimental parameters given by the legend.}
\label{vzdists}
\end{center}
\end{figure}

\begin{figure}[tb]
\begin{center}
\includegraphics[width=8.6cm]{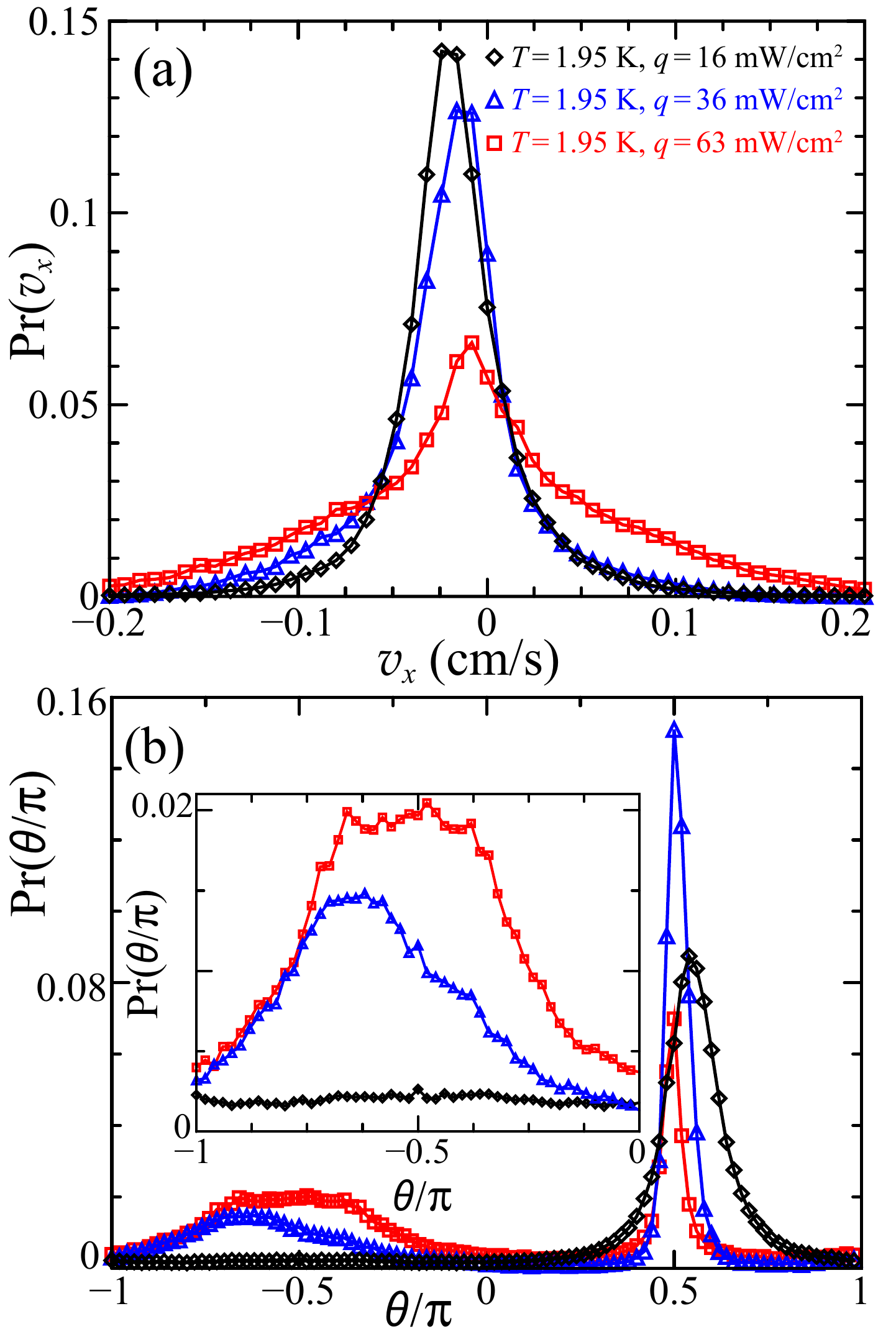}%
\caption{Horizontal velocity and trajectory angle statistics for example counterflow experiments with the parameters given by the legend in (a).  (a) Horizontal velocity $v_x$ distributions.  (b) Distributions of the trajectory angle $\theta \equiv \arctan (v_z/v_x)$, with $\theta/\pi=0.5$ corresponding to trajectories aligned with the direction of the heat flux ($\hat{z}$).  The inset shows only trajectories that move downward ($v_z<0$).}
\label{vx_theta_dists}
\end{center}
\end{figure}

To test these interpretations we compute distributions of the vertical and horizontal velocity components, $v_z$ and $v_x$, which are shown in Figs.\ \ref{vzdists} and \ref{vx_theta_dists}.  The velocities are computed by performing a least squares fit of the form $\mathbf{x}(t)=\mathbf{v}t+\mathbf{x}(0)$ for $0<t<0.1$~s.  In most cases the vertical velocity distributions are bimodal, as expected.  The fraction of downward-moving trajectories increases with: (1) decreasing temperature for constant heat flux and (2) increasing heat flux at constant temperature.  These trends may be attributed to: (1) the increase in vortex line-length density that occurs by increasing either $\left | \mathbf{v}_{\mathrm{n}}-\mathbf{v}_{\mathrm{s}} \right |$ or $q$ and (2) the increase in the particle trapping force with decreasing temperature.  It is important to note, however, that for a given temperature particles will cease to trap in the vortex tangle above a given heat flux (e.g. black circles in Fig.\ \ref{vzdists}(a)).

The distributions of $v_x$ also exhibit a general trend; $\mathrm{Pr}(v_x)$ broadens with increasing fraction of downward-moving trajectories, as shown in Fig.\ \ref{vx_theta_dists}(a). Such behavior is exhibited by the trajectories shown in Fig.\ \ref{cf_traj} consisting of upward-moving trajectories that are vertical with $v_x$ near zero for all times, whereas particles trapped in the vortex tangle move erratically in the $x$-direction, producing a broad distribution of horizontal velocities.  This distinction is further illuminated by the distributions of the trajectory angle $\theta \equiv \arctan (v_z/v_x)$ shown in Fig.\ \ref{vx_theta_dists}(b).  In every experiment $\mathrm{Pr}(\theta/\pi)$ has a sharp peak near $\theta/\pi=0.5$, which corresponds to trajectories aligned with the direction of the heat flux, and thereby aligned with $\mathbf{v}_{\mathrm{n}}$.  In cases with a significant fraction of downward-moving trajectories, a peak also develops at $\theta/\pi=-0.5$, as shown in the inset of Fig.\ \ref{vx_theta_dists}(b).  These secondary peaks are often flat and centered at $\theta/\pi=-0.5$.  We believe that these observations can be explained by mutual friction \cite{vinen57}, which is one source of the horizontal velocity component for the vortex tangle.  However, to do so, the geometry of the vortices needs to be specified and the subject may, therefore, be more appropriate for numerical simulations.

\begin{figure}[tb]
\begin{center}
\includegraphics[width=8.6cm]{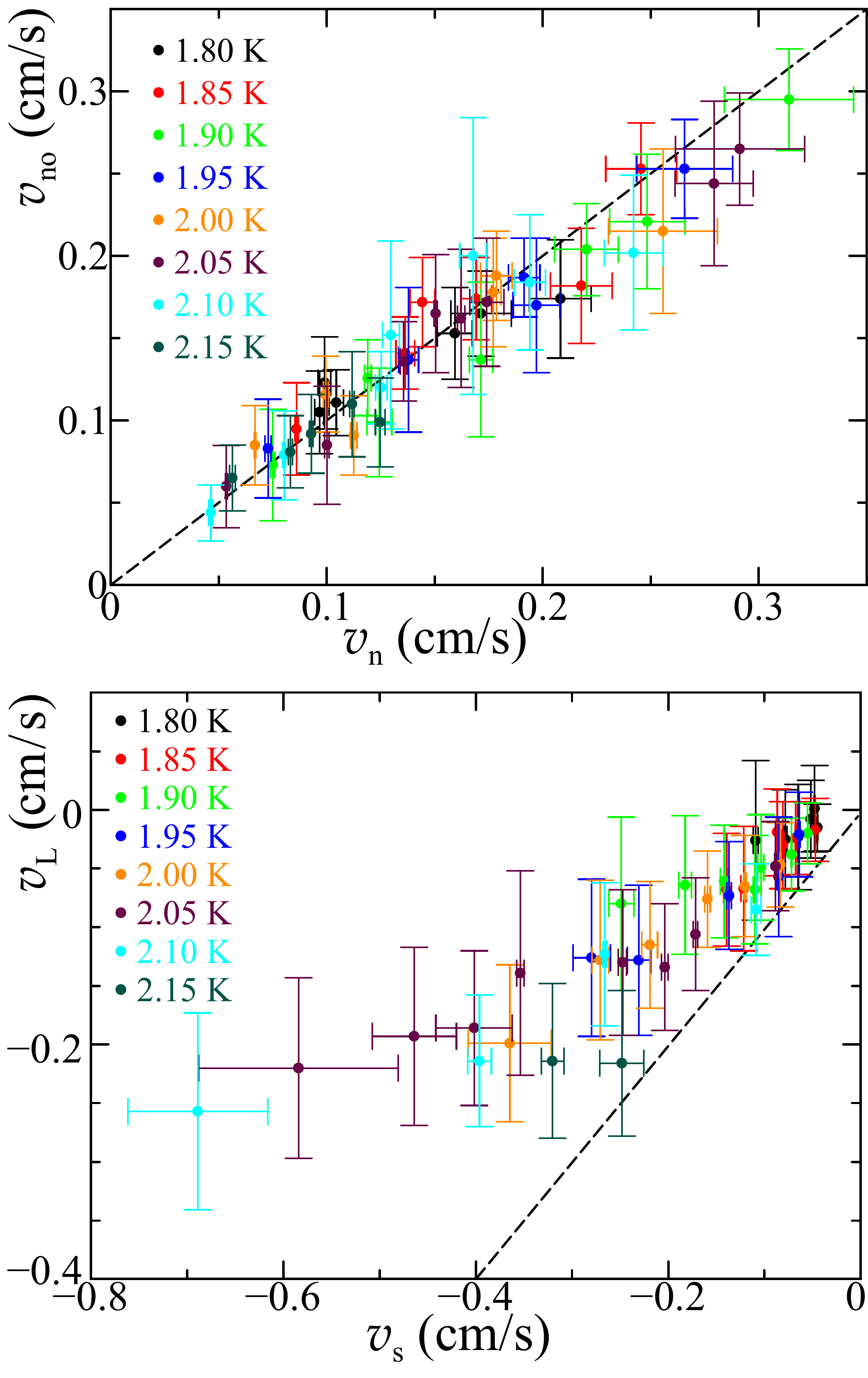}%
\caption{(a) Observed normal fluid velocity $v_{\mathrm{no}}$ as a function of the calculated normal fluid velocity $v_{\mathrm{n}}$ given by eq.~(\ref{vn}). The dashed line corresponds to the prediction $v_{\mathrm{no}}=v_{\mathrm{n}}$.  The horizontal error bars in both plots are due to temperature variations over the course of the run while the vertical error bars are given by the standard deviations computed by the Gaussian fits to each peak. (b) Vertical velocity of the vortex tangle $v_{\mathrm{L}}$ as a function of the calculated superfluid velocity $v_{\mathrm{s}}$ given by eq.~(\ref{vs}).  The dashed line corresponds to $v_{\mathrm{L}}=v_{\mathrm{s}}$.}
\label{vplots}
\end{center}
\end{figure}

To compare with eq.~(\ref{vn}), the observed normal fluid velocity $v_{\mathrm{no}}$ is computed for each experiment by fitting a Gaussian to the vertical velocity distribution function $\mathrm{Pr}(v_z)$ for $v_z>0$.  Similarly, the $\hat{z}$-component of the vortex tangle velocity $v_{\mathrm{L}}$ is computed by fitting a Gaussian to $\mathrm{Pr}(v_z)$ for $v_z<0$.  If there is no peak for $v_z<0$, as is the case for the black circles in Fig.\ \ref{vzdists}(a), we do not compute $v_{\mathrm{L}}$.  The observed normal fluid velocities are shown in Fig.\ \ref{vplots}(a).  For all experimental parameters $v_{\mathrm{no}}$ agrees with $v_{\mathrm{n}}$, giving a quantitative confirmation of eq.~(\ref{vn}).  The values of $v_{\mathrm{L}}$, though, are systematically less negative than $v_{\mathrm{s}}$, particularly for higher heat fluxes and temperatures, as shown in Fig.\ \ref{vplots}(b).  This may be due to mutual friction, but we cannot comment further on the observed values of $v_{\mathrm{L}}$.

\subsection{Discussion}
The results of these studies do not agree with the previous experiments of Zhang and van Sciver as discussed in ref.~2.  Mostly notably, we find that tracer particles that are not trapped by quantized vortices closely follow the calculated normal fluid velocity given by eq.~(\ref{vn}), whereas in ref.~2 the particle velocity $v_{p,a}$ was measured to be approximately $0.5v_n$ (see Fig.\ 1 in ref.~2) independent of $T$.  Several important features must be considered when comparing these studies.  The velocities in ref.~2 were determined using particle image velocimetry (PIV), which computes the velocity using cross-correlation of sub-images containing many tracer particles and is designed for smoothly varying velocity fields.  However, as evidenced in Fig.\ \ref{cf_traj}, the velocity field as computed from the particle trajectories is not smoothly varying owing to the interaction with quantized vortices.  By co-mingling tracers that are moving primarily under the influence of Stokes drag with those trapped in the vortex tangle when measuring the velocity, it is very difficult to disentangle the effects of the quantized vortices on the tracer particles.  The heat fluxes used in the two sets of experiments (110 to 1370 mW/cm$^2$ in ref.~2 and 13 to 91 mW/cm$^2$ here) are disjoint but adjacent.  Even so, there is a discrepancy between our data and those of Zhang and van Sciver.  This discrepancy may be due to the difference in analysis technique.  In particular, Zhang and Van Sciver's analysis did not allow for the influence of quantized vortices on the trajectories of \textit{individual} tracer particles, as we have done.

The theoretical explanation given in ref.~12 by Sergeev \textit{et al.} for the results of Zhang and Van Sciver also appears to be different from ours.  According to their calculations, we should have also seen significant deviations from eq.~(\ref{vn}).  The underlying assumption in their calculation that seems to break down is that every particle is affected by the quantized vortices.  As we have shown in Fig.\ \ref{cf_traj}, a fraction of the particles move on vertical trajectories unaffected by quantized vortices, whereas other trajectories are dominated by their motion.  Lastly, the temperature dependence in both our experiments and previous work must be better understood.  In our experiments, there is a very clear temperature dependence of the effects of the quantized vortices on the particle motions as evidenced by the varying fraction of downward-moving trajectories.  However, in both the experiments of Zhang and Van Sciver and the theoretical explanation given by Sergeev \textit{et al.}, the claim is made that the disparity between the particle velocities and $v_{\mathrm{n}}$ is independent of temperature.  Clearly the motion of tracer particles in thermal counterflow requires further attention.

\section{Conclusions} \label{conclusions}
We have presented a discussion of initial experiments using solid hydrogen tracers to probe the dynamics of superfluid $^4$He.  The tracers can move with the normal fluid under the action of Stokes drag or be trapped or scattered by the quantized vortices.  The response of the particles depends greatly on temperature, particle and vortex characteristics, and flow properties.  These observations serve as only a beginning of a detailed study of the dynamics of superfluid $^4$He on mesoscopic length scales (i.e., between the vortex core size and the intervortex spacing).  Several issues remain unsolved and require further exploration experimentally, numerically, and theoretically, especially those involving the crossover between trapping and scattering as well as interparticle forces along the quantized vortices.

\section*{Acknowledgment}
We thank Michael Fisher, Makoto Tsubota, Nigel Goldenfeld, Russell Donnelly, and Christopher Lobb for helpful discussions and Gregory Bewley for prior collaboration, and are grateful for the financial support of NSF-DMR and the Center for Nanophysics and Materials at the University of Maryland.

%\bibliography{JPSJ}

\begin{thebibliography}{99}
\bibitem{zhang04} T. Zhang and S.~W. van Sciver: Phys. Fluids \textbf{16} (2004) L99.
\bibitem{zhang05a} T. Zhang and S.~W. van Sciver: J. Low Temp. Phys. \textbf{138} (2005) 865.
\bibitem{zhang05b} T. Zhang and S.~W. van Sciver: Nature Phys. \textbf{1} (2005) 36.
\bibitem{vansciver07} S.~W. van Sciver, S. Fuzier and T. Xu: J. Low Temp. Phys. \textbf{148} (2007) 225.
\bibitem{xu07} T. Xu and S.~W. van Sciver: Phys. Fluids \textbf{19} (2007) 071703.
\bibitem{bewley06} G.~P.~Bewley, D.~P.~Lathrop and K.~R.~Sreenivasan: Nature \textbf{441} (2006) 588.
\bibitem{bewley07} G.~P.~Bewley, M.~S.~Paoletti, D.~P.~Lathrop and K.~R.~Sreenivasan: \textit{Proc. IUTAM Symp.}, ed. Y. Kaneda, Nagoya, Japan, 2007.
\bibitem{bewley08} G.~P.~Bewley, M. S. Paoletti, K. R. Sreenivasan and D.~P. Lathrop: to be published in Proc. Natl. Acad. Sci. U.S.A
\bibitem{paoletti08} M.~S. Paoletti, M.~E. Fisher, K.~R. Sreenivasan and D.~P. Lathrop: to be published in Phys. Rev. Lett.
\bibitem{parks66} P.~E.~ Parks and R.~J.~Donnelly: Phys. Rev. Lett. \textbf{16} (1966) 45.
\bibitem{poole05} D.~R.~Poole, C.~F.~Barenghi, Y.~A.~Sergeev and W.~F.~Vinen: Phys. Rev. B \textbf{71} (2005) 064514.
\bibitem{sergeev06} Y.~A.~Sergeev, C.~F.~Barenghi and D.~Kivotides: Phys. Rev. B \textbf{74} (2006) 184506.
\bibitem{barenghi07b} C.~F. Barenghi, D. Kivotides and Y.~A. Sergeev: J. Low Temp. Phys. \textbf{148} (2007) 293.
\bibitem{kivotides07} D.~Kivotides, C.~F. Barenghi and Y.~A. Sergeev: Phys. Rev. B \textbf{75} (2007) 212502.
\bibitem{kivotides08} D.~Kivotides, C.~F. Barenghi and Y.~A. Sergeev: Phys. Rev. B \textbf{77} (2008) 014527.
\bibitem{bewleythesis} G.~P. Bewley: PhD Thesis, Department of Mechanical Engineering, Yale University, New Haven, CT USA (2006).
\bibitem{landau41} L.~Landau: Phys. Rev. \textbf{60} (1941) 356.
\bibitem{tisza47} L.~Tisza: Phys. Rev. \textbf{72} (1947) 838.
\bibitem{weeks} We thank Eric Weeks and John Crocker for the particle-tracking algorithm.
\bibitem{landau} L.~D. Landau and E.~M. Lifshitz: \textit{Fluid Mechanics} (Pergammon Press, New York, 1987) 2nd ed., eq.~(137.1).
\bibitem{vinen57} W.~F.~ Vinen: Proc. Roy. Soc. A \textbf{242} (1957) 493.
\end{thebibliography}

\clearpage

\end{document}